\documentclass[letter]{aa}

\usepackage{graphicx}
\usepackage{xcolor,colortbl,ulem}

\usepackage{txfonts}

\usepackage{xcolor}
\definecolor{darkblue}{rgb}{0.0, 0.0, 0.55}
\usepackage[colorlinks=true, allcolors=darkblue]{hyperref}
\usepackage{orcidlink}

\begin{document}

   \title{Towards systematic searches for LISA white dwarf binaries \\ with multiband photometry}

   \author{A. Perego \orcidlink{0009-0001-0670-2738} \inst{1,2}, A. Lamberts \orcidlink{0000-0001-8740-0127} \inst{2,1}, M. Schultheis \orcidlink{0000-0002-6590-1657} \inst{2},  N. Christensen \orcidlink{0000-0002-6870-4202} \inst{1}
          }

   \institute{Université Côte d’Azur, Observatoire de la Côte d’Azur, Laboratoire Artemis, CNRS, Bd de l’Observatoire, 06300 Nice, France
         \and
             Université Côte d’Azur, Observatoire de la Côte d’Azur, Laboratoire Lagrange, CNRS, Bd de l’Observatoire, 06300 Nice, France
             }

   \date{Received July 18, 2025; accepted August 19, 2025}

  \abstract
   {Ultra-compact double white dwarfs (DWDs) represent key targets for multi-messenger astrophysics, as they may be observed both through gravitational waves and the electromagnetic (EM) spectrum. The future Laser Interferometer Space Antenna (LISA) will detect thousands of these systems, and they are predicted to be the most numerous science targets of the mission.}
   {We develop a strategy to identify LISA source candidates in multiband photometric surveys.}
   {We constructed a synthetic EM catalogue of white dwarf (WD) detections based on a population synthesis code combined with a semi-analytical model of the Milky Way and a consistent cooling model for the evolution. We compared sources in the LISA band with other WD observations in magnitude-colour and colour-colour plots.}
   {From a full sky survey with $u \le$24.5, we find that 57$\%$ of the sources  in the LISA band occupy a specific region in colour-colour diagrams. Inside this area, we find that $\sim 63\%$ (6.5 $\times 10^4$) of EM observations are LISA candidates,  $\sim 31\%$ ($ 3.2 \times 10^4$) are DWDs slightly outside the LISA frequency range, and only a small contamination comes from single WDs and wider binaries.}
   {We find that the colour distributions of close DWDs represent a powerful tool to distinguish potential LISA sources from the broader WD population. This is an avenue to select candidates for further follow-up and identification.}


\titlerunning{LISA white dwarf binaries}

   \maketitle

\section{Introduction}

Low-mass stars constitute the bulk of the stellar population of the Milky Way (MW). At the end of their lives, they evolve into white dwarfs (WDs), dense stellar remnants supported by electron degeneracy pressure. A notable fraction of these objects can be found in binary systems, and of particular interest are double white dwarfs (DWDs) with orbital periods lower than $\sim$ 90 minutes \citep{Kupfer_2024}, which are slowly inspiralling due to gravitational wave (GW) emission \citep{Nelemans_2001}. This makes DWDs the most numerous class of sources detectable by the upcoming {Laser Interferometer Space Antenna} \citep[LISA;][]{LISA_redbook} expected to launch in 2035. Thanks to its frequency sensitivity in the millihertz band, this GW space-based observatory is predicted to detect thousands of DWDs, opening the door to a complete view of close DWDs in the whole MW \citep{LISA_astro, Korol_2017, Korol_2018}.

Despite their relatively low luminosity ($u_{abs}$ $\sim12$ for a 1-Gyr-old WD), several electromagnetic (EM) large-area surveys of WDs have been achieved, such as the spectroscopic Sloan Digital Sky Survey \citep[SDSS;][]{Kepler_2016} and the \textit{Gaia} astrometric and spectroscopic survey \citep{Fusillo_2021}. Binary systems at wide separations can be identified through astrometric measurements \citep{El-Badry_2021}, whereas close binaries present greater challenges because their components cannot be individually resolved. A deeper spectroscopic analysis may uncover single or double-lined binaries, as done by the Extremely Low Mass (ELM) survey \citep{Brown_2020}. Alternatively, light curves from high-cadence photometric surveys, such as the \textit{Zwicky} Transient Factory (ZTF; \citealt{ZTF}), may reveal eclipse signatures or ellipsoidal variations \citep{Burdge_2020, vanRoestel_2025}. Lastly, several compact WD binaries have been detected as AM CVn systems in which a WD accretes material from a low-mass donor star, sometimes showing outbursts or short-period X-ray variations \citep{Ramsay_2018, Green_2024}.

Studying tight DWDs is crucial for understanding stellar and binary evolution processes such as mass transfer episodes,  including common envelope phases. The study of these systems could be substantially enhanced by combining GW and EM measurements, which yield far more information than either method individually. EM data can improve the accuracy of GW parameters (e.g. sky localisation, inclination) while also allowing for the measurement of individual component masses \citep{Shah_2014} and/or radii. If changes in the orbital frequency can be measured by both EM and GW, one can infer the presence of tidal interactions \citep{Shah_2014_b, Leslie_2025}, and this would aid in determining whether the binary will ultimately merge or not and whether it could be a type Ia supernova progenitor. In addition to these scientific objectives, a reliable set of well-characterised EM binaries would also provide significant benefits for LISA data analysis. Their signals could be used to calibrate the instrument, validate the analysis pipeline, and test the data quality \citep{Savalle_2022,Littenberg_24_verification_binaries}.

With a scheduled launch of 2035, now is the time to prepare the systematic search for these systems. In preparation for LISA, our goal is to expand the list of LISA candidate systems that exhibit detectable EM signatures using their characteristic nature of ultra-compact DWDs. In this letter, we provide a strategy for their identification in photometric data, paving the way for subsequent follow-up studies. Section \ref{sec:methods} outlines the procedure used to construct a simulated sample of EM observations of WDs, while in Section \ref{sec:results} we present the resulting density distribution in magnitudes and colours, which is further discussed in Section~\ref{sec:discussion}.

\section{Methods}
\label{sec:methods}

   In this letter we investigate the EM signatures of WDs in the $u$, $g$, and $r$ photometric bands by building a synthetic WD catalogue including single WDs, DWDs, and white dwarf-main sequence (WDMS) binaries. When two WDs are in a binary system, their fluxes blend together, mimicking a single source in photometric surveys. Moreover, we considered WDs paired with a faint main sequence companion since the observed magnitudes are dominated by the brighter WD, and their identification as a WDMS binary is more challenging \citep{Rebassa_2021}, leading to the contamination of the sample.
   
   We employed the binary population synthesis code COSMIC \citep{Breivik_2020} to generate a population of WDs in the MW, either as single stars or as DWDs or WDMS binaries. The systems were initialised with multi-dimensional parameter distributions following \citet{Moe_2017}, while for the binary star evolution, we used the default parameters of COSMIC.\footnote{\url{https://github.com/COSMIC-PopSynth/COSMIC/blob/master/examples/Params.ini}} We maintained this configuration for all the runs and varied the stellar metallicity across 25 distinct values logarithmically spaced between $1.07\times 10^{-4}$ to $2.69\times 10^{-2}$. The simulated population provides the orbital period and formation time of DWDs and WDMS binaries along with the intrinsic properties of individual stars, such as mass and progenitor age. Particularly relevant is the core composition of the WDs. In general, the majority of them have a carbon--oxygen (CO) core, with a small fraction of oxygen--neon (ONe) WDs resulting from the evolution of more massive stars. In addition, helium-core (He) WDs can originate from binary stars whose envelope has been stripped away. All of these physical features strongly influence the subsequent evolution up to the present time and the resulting EM signature.

    We assigned coordinates to each system following a three-component model of the MW, with a bulge, halo, and disc. We referred to the spatial distribution of  \citet{Nelemans_2001} for the disc, defining a scale height of 300 pc; to \citet{Ruiter_2009} for the halo; and to \citet{Korol_2017} for the bulge. The density of the halo was normalised with a mass fraction of WDs of 15\%  of the total baryonic mass of $10^9 M_{\odot}$ \citep{Ruiter_2009}. The bulge was assumed to contain half the mass of the disc. For the disc and the bulge, we implemented an overall star formation history derived from a cosmological hydrodynamic simulation of a MW-like galaxy, employed also in \citet{Lamberts_2019}, that accounts for the evolution in time of the mean metallicity. We assumed a halo composed of stars born in the first $\sim$ 1.5 Gyr with Z $\lesssim 0.0008$.
   
   We let the DWDs  evolve  through GW emission \citep{Peters_1963}, assuming circular orbits, and we discarded the ones that coalesced before the present time. We neglected DWD interactions but remark that strong tidal torques can significantly modify the final population by slowing the evolution and causing binaries to outspiral \citep{Toubiana_2024}. The final population of systems in the MW at the present time comprises $\sim 8.7\times 10^8$ DWDs, $\sim 2.5\times 10^9$ WDMS binaries, and $\sim 5.3\times 10^9$ single WDs.
   
    We now focus on the evolution of each  single WD. As WDs do not sustain nuclear reactions in the core, they only cool down due to various physical mechanisms. The WD cooling time is highly multi-faceted and primarily governed by the core composition, the mass, and the luminosity \citep{Althaus_2010}. Several cooling curves have been computed up to now based on different models and assumptions. In our work, we employ the cooling sequences for pure H atmosphere WDs from \citet{LaPlata_tracks}.\footnote{\url{https://evolgroup.fcaglp.unlp.edu.ar/modelos.html}} These numerical tracks have been computed by interpolating in the mass range 0.2 - 1.3 $M_{\odot}$ the full evolutionary calculations of \citet{Althaus_2013} for He WDs, those of \citet{Camisassa_2016} for CO WDs, and thoes of \citet{Camisassa_2019} for ONe WDs. The tables also provide absolute magnitude values on various photometric systems based on the atmosphere calculations of \citet{Koester_2010}.

    \begin{figure}
    \centering
    \includegraphics[width=\hsize]{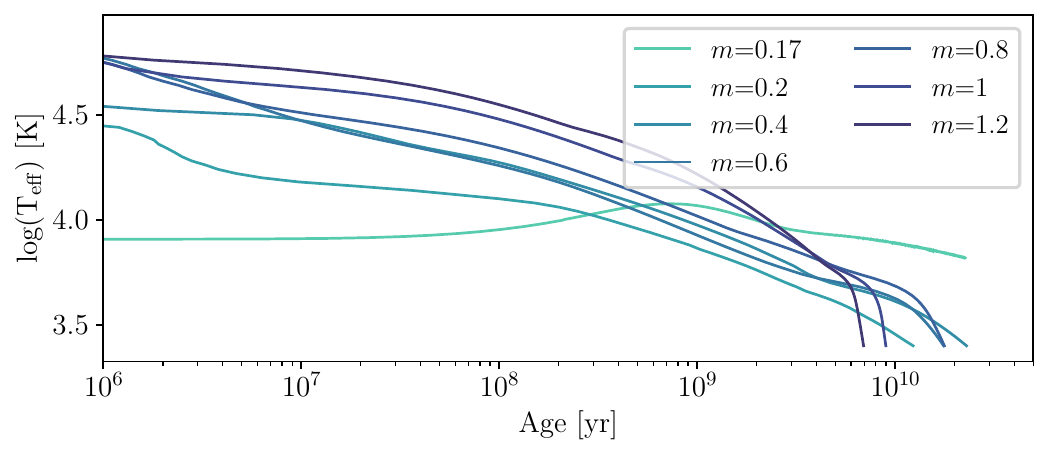}
    \caption{Effective temperature of the WD as a function of its age for different mass values. See the main text for the derivation of the curves.
         }
    \label{fig:cooling_curves}
    \end{figure}

    Cooling models for He WDs show an evident age dichotomy occurring at 0.18 $M_{\odot}$, which is due to the presence or lack of diffusion-induced CNO thermonuclear flashes in the first stages of cooling. The interpolated curves discussed previously do not account for this feature, as they range until 0.2 $M_{\odot}$. Therefore, for extreme low-mass WDs, we used the full cooling computations from \citet{Althaus_2013} to infer their temperature. However, as these calculations do not provide magnitudes, we still relied on the previous interpolated sequences of 0.2 $M_{\odot}$ for them, scaling the flux by the radius as $R^2 / R_{0.2 M_{\odot}}^2$. Therefore there might be some discrepancy in treatment of these systems, but we remark that the fraction of WDs with a mass below 0.18 $M_{\odot}$ is  $\sim 10^{-3}$ of the final dataset of observable WDs, limiting the impact of our choice. Figure \ref{fig:cooling_curves} displays the final cooling sequences for different masses. By interpolating these curves in a continuous range of mass and age, we extracted the WD properties at the present time for all the WDs in our population. We remark that we do not account for tidal heating, which can contribute to the WD luminosity in ultra-short-period binaries, making them brighter and potentially easier to detect.

    The main sequence stars were approximated with a black body spectrum based on the effective temperature and stellar radius provided by COSMIC, which adopts the detailed single-star evolution formulae of \citet{Hurley_2000}. We applied the filter response curve for SDSS \citep{Doi_2010} to obtain the absolute magnitudes in the $u$, $g$, and $r$ bands. Since we are interested in WD-dominated WDMS binaries, we included in our population only the binaries in which the WD is at least 150$\%$ more luminous in absolute magnitude than the companion main sequence star. For these systems, together with all the DWDs, we computed the total absolute magnitude by summing the fluxes of the two components in different bands.
    
Finally, we obtained the observed emission of each WD by computing the apparent magnitude, $u_{\rm app}$, as
    \begin{equation}
        u_{\rm app} = u_{\rm abs} -5 +5\log{d} +A_{u},
    \end{equation}
where $u_{\rm abs}$ is the absolute magnitude in the $u$ filter, $d$ is the distance from the WD in parsecs, and $A_{u}$ is the Galactic extinction in the $u$ band. In order to estimate the latter, we relied on the three-dimensional map of MW dust reddening generated by \citet{Green_2019}. As the map covers only three-quarters of the sky, we completed the sky coverage by incorporating the all-sky extinction map provided by \citet{full_dust_map}. The latter is valid only up to 2.8 kpc in distance, which we find to be sufficient for most of the sources, as shown later. We used the coefficients from \citet{Schlafly_2011} to convert the extinction values in different band passes.

\section{Results}
\label{sec:results}

    As WDs are typically hot astrophysical objects, the $u$ filter becomes particularly useful for studying their emission together with $g$ and $r$ filters to have colour combinations. We constructed a mock dataset of all-sky EM observations in these bands by assuming an observed magnitude limit of 24.54 in the $u$ and $r$ bands and 25.07 in the $g$ filter, consistent with the capabilities of instruments such as MegaCam of the Canada-France-Hawaii Telescope (CFHT).

\begin{figure}
        \centering
        \includegraphics[width=\hsize]{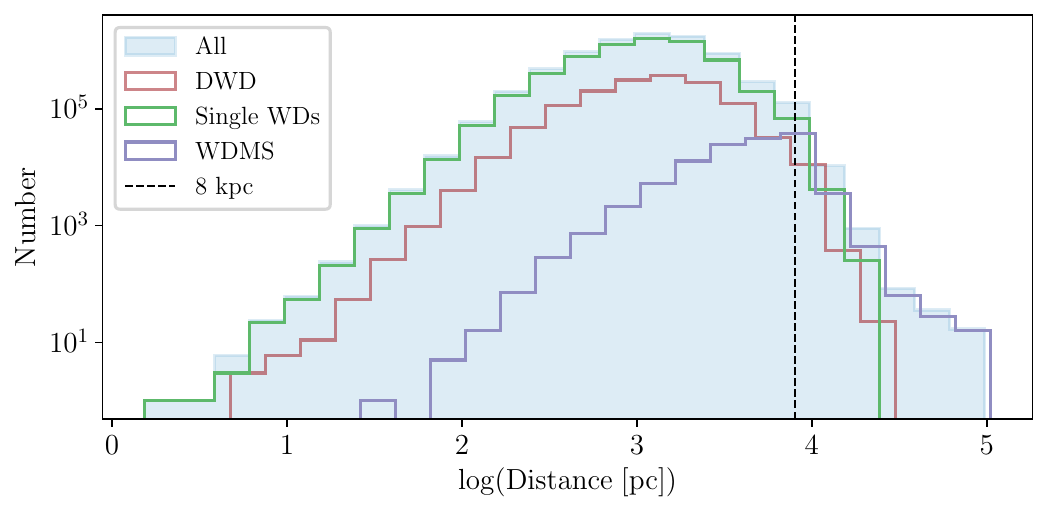}
        \caption{Number of observable systems below the magnitude limit as a function of distance with the different contributions from DWDs, WDMS binaries, and single WDs.
             }
        \label{fig:dist_hist}
        \end{figure}

    Within these magnitude limits, the simulated EM catalogue contains $\sim 1.51\times 10^6$ DWDs, $\sim 1.19\times 10^5$ WDMS binaries, and $\sim 6.65 \times 10^6$ single WDs. In Figure \ref{fig:dist_hist} we show the histogram of the distance for all the observable sources in different subclasses. The peak of the distribution lies around 1 kpc, except for the WDMS binaries since these have already gone through a selection in absolute magnitude of the brightest WDs, which can be seen at greater distances and thus altering the distance profile.

\begin{figure}
        \centering
        \includegraphics[width=\hsize]{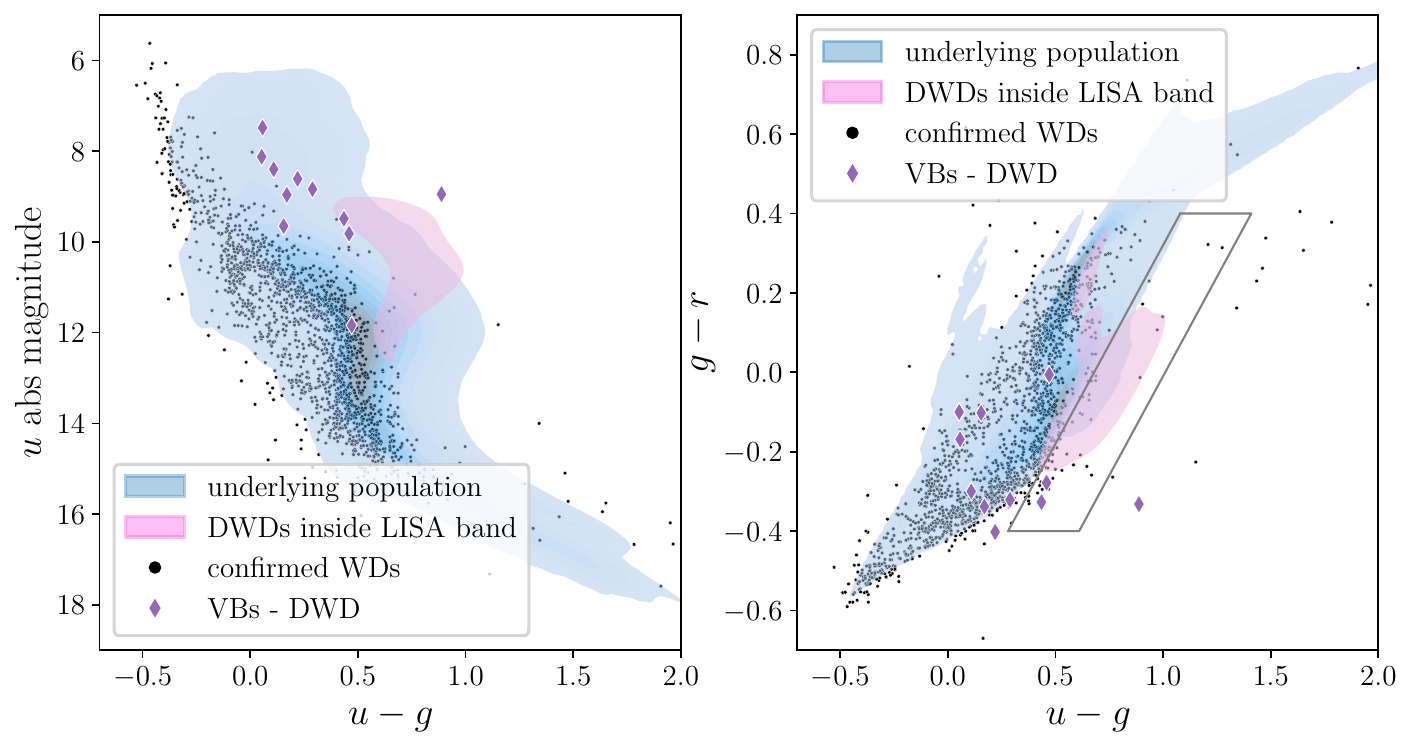}
        \caption{Density distribution in magnitude-colour (left) and colour-colour space (right) of our synthetic catalogue highlighting potential LISA sources (magenta) with respect to the rest of the remaining systems (blue). The black points are confirmed WD observations, while purple markers are detached DWD VBs. We note that several VBs fall directly in the magenta area when reddening is removed and a more stringent magnitude limit is applied. On the right, the grey lines define the colour cuts used in our analysis. 
             }
        \label{fig:u_ug_plot}
        \end{figure}

    We started investigating the properties of our synthetic catalogue by looking at the absolute $u$ magnitude and the $u$--$g$ colour. In the left panel of Figure \ref{fig:u_ug_plot}, we have plotted the density distribution of the sources, differentiating between possible LISA sources (in magenta) and the rest of the sample (in blue). We considered sources with GW emission between $[10^{-4}-0.1]$ Hz to be possible LISA sources ($\sim 1.1 \times 10^5$ systems in our catalogue). The GW frequency of a circular binary system is twice the orbital frequency for a monochromatic signal. The total population also contains DWDs at frequencies lower than the LISA band, single WDs, and WDs in WDMS binaries. It is remarkable how the high frequency DWDs stand out from the broader population. This is because most of them have undergone binary interactions and stellar stripping and are characterised by lower mass values and larger radii, which directly lead to brighter observed emissions. In addition, LISA sources have been found to be younger WDs and therefore are intrinsically more luminous \citep{Lamberts_2019}. These distinctive traits can facilitate the identification of LISA candidates among all the EM observations, enabling a targeted analysis of a particular area in the magnitude–colour plot.
    
    We validated the simulated catalogue with WD candidates from \textit{Gaia} Data Release 3 \citep{Fusillo_2021}. We selected a highly reliable sample of 41820 WDs that have additional SDSS spectroscopy. This sample contains \textit{Gaia} photometry (Early Data Release 3) as well as SDSS photometry. We refer to \citet{Fusillo_2021} for a more detailed description of the catalogue. The observations in the SDSS system are displayed in Figure \ref{fig:u_ug_plot} as black dots. Their location in the $u$ versus $u$--$g$ diagram aligns with our synthetic population, supporting the validity of our theoretical framework.

    In recent years, a limited sample of very short-period compact binaries has been discovered through EM observations thanks to surveys such as ZTF and ELM. In the framework of the LISA mission, these systems are commonly referred to as verification binaries (VBs), as their GW emission lies securely within the interferometer’s sensitivity band \citep{Kupfer_2024}. We are particularly interested in detached DWDs that are cross-matched with SDSS data in order to have a direct and reliable measurement of the observed magnitudes in the $u$, $g$, and $r$ bands. For each system, the absolute magnitudes were inferred from the available distances and the extinction estimate was computed as explained in Section \ref{sec:methods}. We have plotted the final magnitudes and colours in Figure \ref{fig:u_ug_plot} with purple diamond-shaped markers.

    In the broader context of photometric surveys, where distance measurements are not available, we turned our attention to colour–colour plots, which allowed us to display data based only on their observed magnitudes. The right panel of Figure \ref{fig:u_ug_plot} shows the density distribution of our synthetic population in the $u$--$g$ versus $g$--$r$ colour space, with LISA candidates emphasised as before. By applying specific colour cuts shown in Figure \ref{fig:u_ug_plot} (-0.4 < $g$--$r$ < 0.4 and -1.01 < $g$--$r$ - $u$--$g$ < - 0.68), we found that the major contribution in that region comes from LISA candidates (63\%, $\sim 65000$ systems), with a smaller contribution (31\%, $\sim 32000$ systems) from DWDs at GW frequencies from $\sim 10^{-7}$ Hz to $\sim 10^{-4}$ Hz, which are still relevant when studying the continuum of short-period DWDs. The LISA candidates inside the selected area are $\sim 57\%$ of all the sample without colour selection. Only a minimal fraction comes from single WDs (5\%), and even less is from ultra-wide DWDs (0.1\%), while we have almost no contribution from WDMS binaries, as their colour distribution covers markedly different values. Between the high frequency DWDs, we estimate $\sim 2000$ of them will be detectable by LISA in 10 years of observation time using the \textsc{GBGPU} package \citep{katz_2022_6500434}.

    \section{Discussion and conclusions}
    \label{sec:discussion}

    We have generated a synthetic population of Galactic WDs, both as single stars and in binary systems, by combining the binary population synthesis code COSMIC with a metallicity-dependent star formation history and an analytic model of the MW. We computed magnitudes and colours from full evolutionary models of WDs. From this sample, we constructed a mock EM catalogue by assuming a $u$ magnitude depth of 24.5 and a full sky area while also accounting for dust extinction. We investigated the density distribution in magnitude-colour and colour-colour space, focusing on the $u$, $g$, and $r$ filters. We find that high frequency DWDs, which are potential LISA sources, clearly separate from the underlying population of wider DWDs, single stars, and WDMS binaries. This feature stems from the distinctive properties of ultra-compact DWDs detectable by LISA,, such as lower mass values and smaller ages, which results from their evolutionary history. These traits are likewise exhibited by compact DWDs at marginally lower frequencies, not falling directly within LISA’s detection window. We applied a colour selection to maximise the fraction of ultra-compact DWDs over the whole sample of observations. In this targeted area in the colour-colour diagram, we found that $\sim 63\%$ of all observations come from DWDs in the LISA band, with a smaller fraction from binaries with orbital frequencies down to $10^{-7}$ Hz and a negligible contamination from single systems, wide DWDs, or WDMS binaries. 

    We repeated the same analysis following the DWD model described in Sections 2.3 and 4.1 of \citet{Boileau_2025}. Apart from having different assumptions for the initial parameter distributions and the common envelope phase, it modifies the criterion for the mass transfer stability of COSMIC so that it resembles the one used by the population synthesis code \textsc{SeBa} \citep{Toonen_2012}. This population model yields a total of $\sim 1.8 \times 10^5$ EM detections of DWDs in the LISA band, but their density distribution in the colour-colour plot overlaps more with the underlying population. After the colour selection, we still found that $\sim 57\%$ of the observations in the targeted area come from LISA candidates and $\sim 40\%$ are from DWDs with a GW frequency down to $10^{-7}$ Hz. In this case, the number of LISA candidates in the selected region is $\sim 7.1 \times 10^4$, which corresponds to $\sim 40\%$ of all the sources. Within this fraction, we estimate $\sim 1900$ DWDs will be detectable by LISA in 10 years.
    
    We also assessed the colours and magnitudes of our synthetic sample in relation to real WD observations from \textit{Gaia} and SDSS and found a good agreement despite different magnitude limits. Most of the observed systems have apparent magnitudes below $21$, which is more restrictive than the upper limit adopted for our model. By implementing this threshold in our catalogue, the distributions showed a complete alignment. This consideration is particularly relevant for the VBs, whose sample is characterised by low apparent magnitudes, making it biased and incomplete. In addition, the diverse identification methods resulted in a highly heterogeneous sample, and certain WD properties, such as different atmospheric composition, may not be fully covered by our model. Due to these considerations, the dataset cannot serve as a comprehensive and accurate representation of LISA multi-messenger observations of DWDs. However, it is worth noting that applying the stricter magnitude limit discussed earlier while also removing the reddening brings the magnitude and colour values of the VBs into closer agreement with those of our LISA candidates sample, with respect to the underlying population.
    
    With our work, we emphasise how the distribution of ultra-compact DWDs in magnitude and colour space naturally separate from the underlying population of WD observations. These results pave the way for tailored strategies aimed at identifying potential LISA sources in photometric surveys, such as the forthcoming community survey on the CFHT.
 
    \begin{acknowledgements}
          This project has received financial support from the CNRS  MITI interdisciplinary programs. A.L. and N.C. acknowledge support by the French ANR. The authors thank the CNES for support for this research.
    \end{acknowledgements}

\bibliographystyle{aa}
\bibliography{bibliography} 

\end{document}